\documentclass[aoas,MSNbibl,nameyear,dvips]{arximspdf}
\usepackage{graphicx}
\usepackage{textcomp}
\doi{10.1214/14-AOAS751} 
\volume{8}
\issue{3}
\pubyear{2014}
\firstpage{1416}
\lastpage{1442}
\docsubty{FLA}

\makeatletter
\newcommand{\rrvert}{\vert}
\newcommand{\llvert}{\vert}
\newcommand{\bftheta}{\bolds{\theta}}
\newcommand{\bfvartheta}{\bolds{\vartheta}}
\newcommand{\bfSigma}{\bolds{\Sigma}}
\newcommand{\bfOmega}{\bolds{\Omega}}
\newcommand{\bfLambda}{\bolds{\Lambda}}
\newcommand{\bfx}{\mathbf{x}}
\newcommand{\bfy}{\mathbf{y}}
\newcommand{\bfB}{\mathbf{B}}
\newcommand{\E}{\mathsf{E}}
\newcommand{\diag}{\operatorname{diag}}
\newcommand{\DP}{\mathsf{DP}}
\newcommand{\Gam}{\mathsf{Gam}}
\newcommand{\IGam}{\mathsf{IGam}}
\newcommand{\IWis}{\mathsf{IWis}}
\newcommand{\bet}{\mathsf{Beta}}
\newcommand{\ber}{\mathsf{Ber}}
\newcommand{\PY}{\mathsf{PY}}
\newcommand{\GDP}{\mathsf{GDP}}
\newtheorem{theorem}{Theorem}
\makeatother

\begin{document}
\begin{frontmatter}

\title{Functional clustering in nested designs: Modeling variability in reproductive epidemiology studies}
\runtitle{Functional clustering in nested designs}

\begin{aug}
\author{\fnms{Abel}~\snm{Rodriguez}\corref{}\ead[label=e1]{abel@soe.ucsc.edu}\thanksref{t1}}
\and
\author{\fnms{David B.}~\snm{Dunson}\thanksref{t2}}
\runauthor{A. Rodriguez and D.~B. Dunson}
\affiliation{University of California, Santa Cruz and Duke University}
\address{Department of Applied Mathematics\\
\quad and Statistics\\
University of California, Santa Cruz\\
1156 High Street\\
Mailstop SOE2\\
Santa Cruz, California 95064\\
USA\\
\printead{e1}} 
\address{Department of Statistics\\
\quad and Decision Sciences\\
Duke University\\
Box 90251\\
Durham, North Carolina 27708\\
USA}
\end{aug}
\thankstext{t1}{Supported in part by Grant R01 GM090201-01 from the National Institute of General Medical Sciences of the National Institutes of Health.}
\thankstext{t2}{Supported in part by Grant R01 ES017240-01 from the National Institute of Environmental Health Sciences of the National Institutes of Health.}

\received{\smonth{7} \syear{2010}}
\revised{\smonth{2} \syear{2014}}

%
\begin{abstract}
We discuss functional clustering procedures for nested designs, where
multiple curves are collected for each subject in the study. We start
by considering the application of standard functional clustering tools
to this problem, which leads to groupings based on the average profile
for each subject. After discussing some of the shortcomings of this
approach, we present a mixture model based on a generalization of the
nested Dirichlet process that clusters subjects based on the
distribution of their curves. By using mixtures of generalized
Dirichlet processes, the model induces a much more flexible prior on
the partition structure than other popular model-based clustering
methods, allowing for different rates of introduction of new clusters
as the number of observations increases. The methods are illustrated
using hormone profiles from multiple menstrual cycles collected for
women in the Early Pregnancy Study.
\end{abstract}

%
\end{frontmatter}

\setcounter{footnote}{2}

\section{Introduction}\label{seintro}\label{sec1}

The literature on functional data analysis has seen a spectacular
growth in the last twenty years, showing promise in applications
ranging from genetics [\citet{RaSeKo02,LuLi03,WaZhSe03}] to proteomics
[\citet{RaMa06}], epidemiology [\citet{BiDu05}] and
oceanography [\citet{RoDuGe08b}]. Because functional data are
inherently complex,
functional clustering is useful as an exploratory tool in
characterizing variability among subjects; the resulting clusters can
be used as a predictive tool or simply as a hypothesis-generating
mechanism that can help guide further research. Some examples of
functional clustering methods include \citet{AbCoMaMo03}, who use
B-spline fitting coupled with $k$-means clustering; \citet
{TaKi03}, who
apply $k$-means clustering via the principal points of random
functions; \citet{JaSu03}, who develop methods for sparsely sampled
functional data that employ spline representations; \citet{GaGo05},
where the robust $k$-means method for functional clustering is
developed; \citet{SeWa05}, who use a Fourier representation for the
functions along with $k$-means clustering; \citet{HeHoSt06}, where a
Bayesian hierarchical clustering approach that relies on spline
representations is proposed; \citet{RaMa06}, who build a hierarchical
Bayesian model that employs a Bayesian nonparametric mixture model on
the coefficients of the wavelet representations; and \citet{ChLi07},
where a $k$-centers functional clustering approach is developed that
relies on the Karhunen--Lo{\`e}ve representation of the underlying
stochastic process generating the curves and accounts for both the
means and the modes of variation differentials between clusters.

All of the functional clustering methods described above have been
designed for situations where a single curve is observed for each
subject or experimental condition. Extensions to nested designs where
multiple curves are collected per subject typically assume that
coefficients describing subject-specific curves arise from a common
parametric distribution, and clustering procedures are then applied to
the parameters of this underlying distribution. The result is a
procedure that generates clusters of subjects based on their average
response curve, which is not appropriate in applications in which
subjects vary not only in the average but also in the variability of
the replicate curves. For example, in studies of trajectories in
reproductive hormones that collect data from repeated menstrual cycles,
the average trajectory may provide an inadequate summary of a woman's
reproductive functioning. Some women have regular cycles with little
variability across cycles in the hormone trajectories, while other
women vary substantially across cycles, with a subset of the cycles
having very different trajectory shapes. In fact, one indication of
impending menopause and a decrease in fecundity is an increase in
variability across the cycles. Hence, in forming clusters and
characterizing variability among women and cycles in hormone
trajectories, it is important to be flexible in characterizing both the
mean curve and the distribution about the mean. This situation is not
unique to hormone data, and similar issues arise in analyzing repeated
medical images as well as other applications.

This paper discusses hierarchical Bayes models for clustering nested
functional data in the context of the Early Pregnancy Study (EPS)
[\citet{WiWeOC88}], where progesterone profiles were collected for both
conceptive and nonconceptive women from multiple menstrual cycles. Our
models use spline bases along with mixture priors to create sparse but
flexible representations of the hormone profiles, and can be applied
directly to other basis systems such as wavelets. We start by
introducing a hierarchical random effects model on the spline
coefficients which, along with a generalization of the Dirichlet
process mixture (DPM) prior [\citet{Fe73,Se94,EsWe95}], allows for
mean-response-curve clustering of women, in the spirit of \citet
{RaMa06}. Then, we extend the model to generate distribution-based
clusters using a nested Dirichlet process (NDP) [\citet{RoDuGe08a}].
The resulting model simultaneously clusters both curves and subjects,
allowing us to identify outlier curves within each group of women, as
well as outlying women whose distribution of profiles differs from the
rest. To the best of our knowledge, there is no classical alternative
for this type of distribution-based multilevel clustering.

%
%
\begin{figure}

\includegraphics{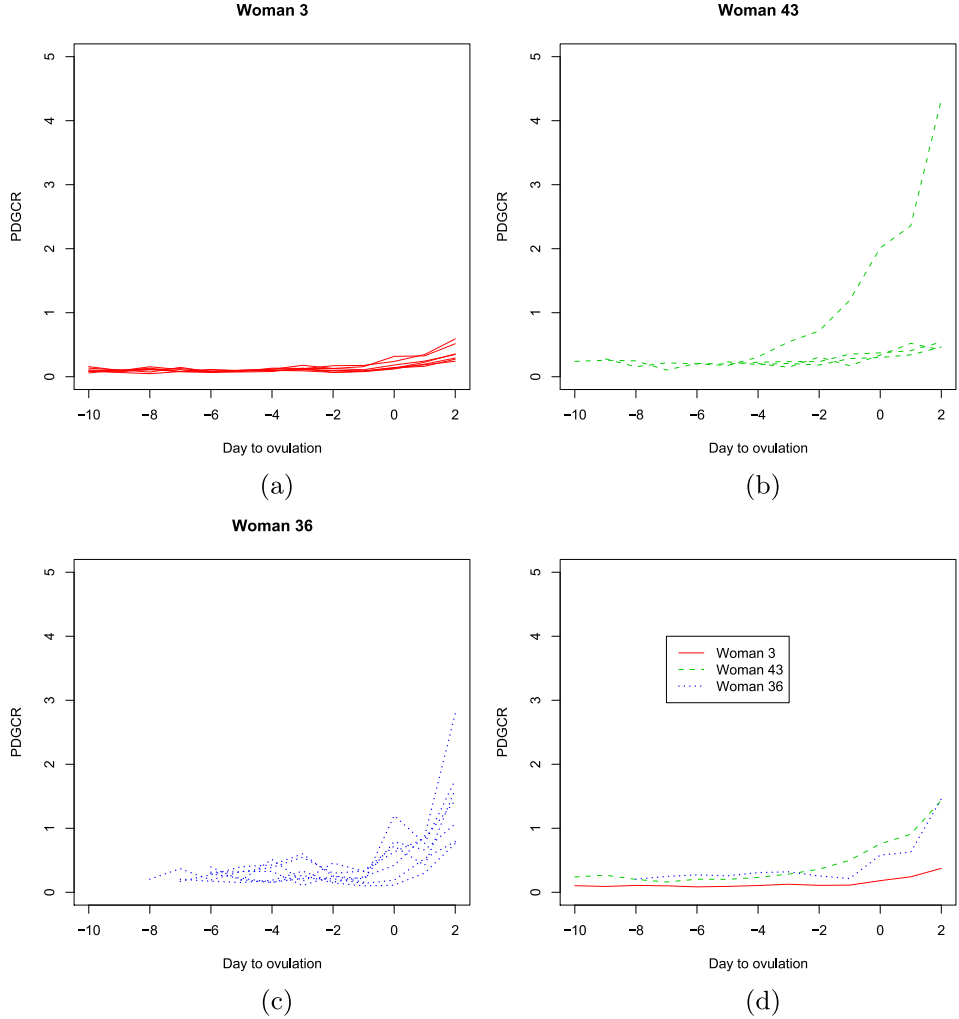}

\caption{Comparison of hormone profiles for three women in the Early Pregnancy Study.
Frames \textup{(a)} to \textup{(c)} show multiple profiles for each woman, while frame
\textup{(d)} shows the average profile for each woman.}\label{fihorpro}
\end{figure}

In order to provide some insight into the challenges associated with
functional clustering in the context of the EPS, consider the hormonal
profiles depicted in Figure~\ref{fihorpro}. Frames (a) to (c) depict
the hormone profiles for 3 women, while frame~(d) shows the mean
profile corresponding to each one of them, obtained by simply averaging
all available observations at a given day within the cycle. When
looking at the mean profiles in (d), women 43 and 36 seem to have very
similar hormonal responses, which are different from those of woman 3.
However, when the individual profiles are considered, it is clear that
most of the cycles of woman 43 look like those of woman 3 and that the
big difference in the means is driven by the single abnormal cycle.
This result, although not surprising if we consider that means are
notoriously nonrobust, suggests that simple approaches that average
profiles over individuals (or, equivalently, use Gaussian distributions
to describe subject-specific variability) might not be the most
appropriate for this type of data.

The use of Bayesian nonparametric mixture models for clustering has a
long history [\citet{MeSi02,QuIg03,LaGr06}] and presents a number of
practical advantages over other model-based clustering techniques.
Nonparametric mixtures induce a probability distribution on the space
of partitions of the data, therefore, we do not need to specify in
advance the number of clusters in the sample. Once updated using the
data, this distribution on partitions allows us to assess variability,
and hence characterize uncertainty in the clustering structure
(including that associated with the estimation of the curves),
providing a more complete picture than classical methods. In this
paper, we work with a generalized Dirichlet process (GDP) first
introduced by \citet{Hj00} and study some of its properties as a
clustering tool. In particular, we show that the GDP generates a richer
prior on data partitions than those induced by popular models such as
the Dirichlet process [\citet{Fe73}] or the two-parameter
Poisson--Dirichlet process [\citet{Pi96}], as it allows for an
asymptotically bounded number of clusters in addition to logarithmic
and power-law rates of growth.

The paper is organized as follows: Section~\ref{semixfunc} reviews the
basics of nonparametric regression and functional clustering, while
Section~\ref{seBNPmixfunc} explores the design of nonparametric
mixture models for functional clustering. Building on these brief
reviews, Section~\ref{sefunclustnested} describes two Bayesian
approaches to functional clustering in nested designs, while
Section~\ref{secompu} describes Markov chain Monte Carlo algorithms
for this problem. An illustration focused on the EPS is presented in
Section~\ref{seillus}. Finally, Section~\ref{seconclusion} presents a
brief discussion and future research directions.



\section{Model-based functional clustering}\label{semixfunc}

To introduce our notation, consider first a simple functional
clustering problem where multiple noisy observations are collected from
functions $f_1, \ldots, f_I$. More specifically, for subjects $i=1,\ldots,I$ and within-subject design points $t=1,\ldots, T_i$,
observations consist of ordered pairs $(x_{it}, y_{it})$, where
%
\begin{eqnarray*}
y_{it} &=& f_i(x_{it}) + \varepsilon_{it},
\qquad\varepsilon_{it} \sim\mathsf{N}\bigl(0, \sigma_i^2
\bigr).
\end{eqnarray*}
For example, in the EPS, $y_{it}$ corresponds to the level of
progesterone in the blood of woman $i$ collected at day $x_{it}$ of the
menstrual cycle, and $f_i$ denotes a smooth trajectory in progesterone
for woman $i$ (initially supposing a single menstrual cycle of data
from each woman), and clusters in $\{ f_i \}_{i=1}^{I}$ could provide
insight into the variability in progesterone curves across women, while
potentially allowing us to identify abnormal or outlying curves.

If all curves are observed at the same covariate levels (i.e., $T_{i} =
T$ and $x_{it} = x_{t}$ for every $i$), a natural approach to
functional clustering is to apply standard clustering methods to the
data vectors, ${\mathbf y}_i = (y_{i1},\ldots, y_{iT})'$. For example, in
the spirit of \citet{RaSi05}, one could apply hierarchical or $k$-means
clustering to the first few principal components [\citet{YeRu01}].
Alternatively, with uneven spacings or missing observations,
nonparametric regression techniques such as kernel regression [e.g.,
see \citet{LiRa04,RaLi04}] could be used to interpolate the value of
the curves to a common grid, and then traditional clustering techniques
could be applied. From a model-based perspective, one could instead
suppose that ${\mathbf y}_i$ is drawn from a mixture of $k$ multivariate
Gaussian distributions, with each Gaussian corresponding to a different
cluster [\citet{FrRa02,YeFrMuRaRu01}]. The number of clusters could
then be selected using the BIC criteria [\citet{FrRa02,Li05}] or a
nonparametric Bayes approach could be used to bypass the need for this
selection, while allowing the number of clusters represented in a
sample of $I$ individuals to increase stochastically with sample size
[\citet{MeSi02}].

A more general approach that allows us to deal with unevenly spaced and
missing observations is to fit a nonparametric model to each curve and
then project all the curves onto a common space. For example, we can
represent the unknown function $f_i$ as a linear combination of
prespecified basis functions $\{ b_k \}_{k=1}^{p}$, that is, we can write
\begin{eqnarray*}
f_i(x_{it}) &=& \theta_{i0} + \sum
_{k=1}^{p} \theta_{ik} b_k(x_{it}),
\end{eqnarray*}
where $\bftheta_i = (\theta_{i0},\theta_{i1}, \ldots, \theta
_{ip})$ are basis coefficients specific to subject $i$, with
variability in these coefficients controlling variability in the curves
$\{ f_i \}_{i=1}^{I}$. A common approach to functional clustering is to
induce clustering of the curves through clustering of the basis
coefficients [\citet{AbCoMaMo03,HeHoSt06}]. Then the methods discussed
above for clustering of the data vectors $\{ {\mathbf y}_i \}_{i=1}^{I}$ in
the balanced design case can essentially be applied directly to the
basis coefficients $\{ \bftheta_i \}_{i=1}^{I}$.

Although the methods apply directly to other choices of basis
functions, our focus will be on splines, which have been previously
used in the context of hormone profiles [\citet{BrRi98,BiDu05}]; given
a set of knots $\tau_1, \ldots, \tau_p$, the $k$th member of the
basis system is defined as 
%
\begin{eqnarray*}
b_k(x) &=& (x - \tau_k)_{+}^{q},
\end{eqnarray*}
where $(\cdot)_{+} = \max\{ \cdot, 0 \}$. Given the knot locations,
inferences on $\bftheta_i$ and $\sigma^2_i$ can be carried out using
standard linear regression tools, however, selecting the number and
location of the nodes $\tau_1,\ldots,\tau_p$ can be a challenging
task. A simple solution is to use a large number of equally spaced
knots, together with a penalty term on the coefficients to prevent
overfitting. From a Bayesian perspective, this penalty term can be
interpreted as a prior on the spline coefficients; for example, the
maximum likelihood estimator (MLE) obtained under an $L^2$ penalty on
the spline coefficients is equivalent to the maximum a posteriori
estimates for a Bayesian model under a normal prior, while the MLE
under an $L^1$ penalty is equivalent to the maximum a posterior
estimate under independent double-exponential priors on the spline coefficients.

Instead of the more traditional Gaussian and double exponential priors,
in this paper we focus on zero-inflated priors, in the spirit of
\citet{SmKo96}. Priors of this type enforce sparsity by zeroing
out some of
the spline coefficients and, by allowing us to select a subset of the
knots, provides adaptive smoothing. In their simpler form,
zero-inflated priors assume that the coefficients are independent from
each other and that
%
%
\begin{eqnarray}
\label{eqzeroinfpr} \theta_{ik} | \gamma, \sigma^{2}_{i}
&\sim& \gamma\mathsf{N}\bigl( 0, \omega_{k} \sigma_i^{2}
\bigr) + (1-\gamma) \delta_{0},\qquad \sigma_i^2
\sim\IGam( \nu_1, \nu_2), 
\end{eqnarray}
where $\delta_{x}$ denotes the degenerate distribution putting all its
mass at $x$, $\omega_{k}$ controls the overdispersion of the
coefficients with respect to the observations and $\gamma$ is the
prior probability that the coefficient $\theta_{ik}$ is different from
zero. In order to incorporate a priori dependence across coefficients,
we can reformulate the hierarchical model by introducing Bernoulli
random variables $\lambda_{i1}, \ldots,\lambda_{ip}$ such that
\begin{eqnarray*}
\bfy_i | \bftheta_i,\sigma^2_i,
\bfLambda_i & \sim& \mathsf{N}\bigl( \bfB(\bfx_i)
\bfLambda_i \bftheta_i, \sigma^2_i
\mathbf{I}\bigr),\qquad\bftheta_i | \sigma_i^2
\sim\mathsf{N}\bigl( \mathbf{0}, \sigma_i^2 \bfOmega
\bigr),
\\
\sigma_i^2 &\sim&\IGam(\nu_1,\nu_2),
\end{eqnarray*}
where $\bfy_i=(y_{i1},\ldots,y_{iT_i})$ and $\bfx_i=(x_{i1},\ldots,x_{iT_i})$ are, respectively, the vectors of responses and covariates
associated with subject $i$, $\mathbf{B}(\bfx_i)$ is the matrix of
basis functions also associated with subject $i$ with entries $[
\mathbf{B}(\bfx_i) ]_{tk} = b_k(x_{it})$, $\bfLambda_i =\diag\{
\lambda_{i1},\ldots, \lambda_{ip} \}$ and $\lambda_i$ equals 1
independently with probability $\gamma$, and $\bfOmega$ is a $p
\times p$ covariance matrix. Note that if $\bfOmega$ is a diagonal
matrix and $[\bfOmega]_{kk} = \omega_k$, we recover the independent
priors in (\ref{eqzeroinfpr}). For the single curve case, choices for~$\bfOmega$ based on the regression matrix $\bfB(\bfx_i)$ are
discussed in \citet{DiGeKa01}, \citet{LiPaMoClBe05} and
\citet{Pa06}.

Although the preceding two-stage approach is simple to implement using
off-the-shelf software, it ignores the uncertainty associated with the
estimation of the basis coefficients while clustering the curves. In
the spirit of \citet{FrRa02}, an alternative that deals with this issue
is to employ a mixture model of the form
%
%
\begin{eqnarray}
\label{eqsimplmodel} \bfy_i | \bigl\{ \bftheta^{*}_{k}
\bigr\}, \bigl\{ \sigma^{*2}_k \bigr\}, \bigl\{\bfLambda
^{*}_k \bigr\} & \sim& \sum_{k=1}^{K}
w_k \mathsf{N} \bigl( \mathbf{B}(\bfx_j)
\bfLambda^{*}_k\bftheta^{*}_{k},
\sigma^{*2}_k \mathbf{I} \bigr),\qquad\sum
_{k=1}^{K} w_k =1,
\end{eqnarray}
where $\bftheta^{*}_k$ is the vector of coefficients associated with
the $k$th cluster, $\bfLambda^{*}_k$ is the diagonal selection matrix
for the $k$th cluster, $\sigma_k^{*2}$ is the observational variance
associated with observations collected in the $k$th cluster, $w_k$ can
be interpreted as the proportion of curves associated with cluster $k$,
and $K$ is the maximum number of clusters in the sample. From a
frequentist perspective, estimation of this model can be performed
using expectation--maximization (EM) algorithms, while selection of the
number of mixture components can be carried out using the BIC.
Alternatively, Bayesian inference can be performed for this model using
Markov chain Monte Carlo (MCMC) algorithms once appropriate priors for
the vector $\mathbf{w} = (w_1,\ldots,w_K)$ and the cluster-specific
parameters $(\bftheta^{*}_k, \bfLambda_k^{*}, \sigma_k^{*2})$ have
been chosen, opening the door to simple procedures for the estimation
of the number of clusters in the sample.

\section{Bayesian nonparametric mixture models for functional data}\label{seBNPmixfunc}

Note that the model in (\ref{eqsimplmodel}) can be rewritten as a
hierarchical model by introducing latent variables $\{ (\bftheta_i,
\sigma_i^2, \bfLambda_i) \}_{i=1}^{I}$ so that
%
%
\begin{eqnarray}\label{eqsimplmodelh}
\bfy_i | \bftheta_i,
\sigma^2_i, \bfLambda_i & \sim& \mathsf{N}
\bigl( \mathbf{B}(\bfx_i)\bfLambda_i
\bftheta_{i}, \sigma^2_i \mathbf{I} \bigr),
\qquad
\bftheta_i, \sigma^2_i,
\bfLambda_i | G \sim G,
\nonumber\\[-8pt]\\[-8pt]
G(\cdot) &=& \sum_{k=1}^{K}
w_k \delta_{(\bftheta^{*}_k, \sigma
^{*2}_k, \bfLambda_k^{*})} (\cdot).\nonumber
\end{eqnarray}
Therefore, specifying a joint prior on $\mathbf{w}$ and $\{ (\bftheta
^{*}_k, \sigma_k^{*2}, \bfLambda^{*}_k)\}_{k=1}^{K}$ is equivalent to
specifying a prior\vspace*{1pt} on the discrete distribution $G$ generating the
latent variables $\{ (\bftheta_i, \sigma_i^2, \bfLambda_i) \}
_{i=1}^{I}$. In this section we discuss strategies to specify a
flexible prior distribution on this mixing distribution in the context
of functional clustering. In particular, we concentrate on
nonparametric specifications for $G$ through the class of
stick-breaking distributions. 

A stick-breaking prior [\citet{IsJa01,OnCa04}] with baseline measure
$G_0$ and precision parameters $\{ a_l \}_{l=1}^{L}$ and $\{ b_l \}
_{l=1}^{L}$ is defined as
%
%
\begin{eqnarray}
G(\cdot) & =& \sum_{k=1}^{K}
w_k \delta_{\bfvartheta_k} (\cdot),\label{eqstikbreak}
\end{eqnarray}
where the atoms $\{\bfvartheta_k \}_{k=1}^{K}$ are independent and
identically distributed samples from $G_0$ and the
weights $\{ w_k \}_{l=1}^{K}$ are constructed as $w_k = u_k \prod_{s<k}
(1-u_s)$, with $\{ u_k
\}_{k=1}^{K}$ another independent and identically distributed sequence
of random variables such
that $u_k \sim\bet(a_k, b_k)$ for $k < K$ and $u_K=1$. For example,
taking $K=\infty$, $a_k=1-a$ and $b_k=b+ka$ for $0 \le a < 1$ and
$b>-a$ yields the two-parameter Poisson--Dirichlet process [\citet
{Pi95,IsJa01}], denoted $\PY(a,b,G_0)$, with the choice $a=0$
resulting in the Dirichlet process [\citet{Fe73,Se94}], denoted
$\DP
(b, G_0)$.
In mixture models such as (\ref{eqsimplmodelh}), $G_0$ acts as the
common prior for the cluster-specific parameters $\{\bfvartheta_k\}
_{k=1}^K$, while the sequences $\{ a_k \}_{k=1}^{K}$ and $\{ b_k \}
_{k=1}^{K}$ control the a priori expected number and size of the clusters.

The main advantage of nonparametric mixture models such as the
Poisson--Dirichlet process as a clustering tool
is that they allow for automatic inferences on the number of components
in the mixture. Indeed, these models induce a prior probability on all
possible partitions of the set of observations, which is updated based
on the information contained in the data. However, Poisson--Dirichlet
processes have two properties that might be unappealing in our EPS
application; first, they place a relatively large probability on
partitions that include many small clusters and, second, they imply
that the number of clusters will tend to grow logarithmically (if
$a=0$) or as a power law (if $a>0$) as more observations are included
in the data set. However, priors that favor \hyperref[sec1]{Introduction} of increasing
numbers of clusters without bound as the number of subjects increase
have some disadvantages in terms of interpretability and sparsity in
characterizing high-dimensional data. For example, in applying DP
mixture models for clustering of the progesterone curves in EPS,
\citet{BiDu05} obtained approximately 32 different clusters,
with half of these clusters singletons. Many of the clusters appeared
similar, and it may be that this large number of clusters was partly an
artifact of the DP prior. \citet{Du09} proposed a local partition
process prior to reduce dimensionality in characterizing the curves,
but this method does not produce easily interpretable functional
clusters. Hence, it is appealing to use a more flexible global
clustering prior that allows the number of clusters to instead converge
to a finite constant.

With this motivation, we focus on the generalized Dirichlet process
(GDP) introduced by \citet{Hj00}, denoted $\GDP(a,b,G_0)$. The GDP
corresponds to a stick-breaking prior with $K=\infty$, $a_k=a$ and
$b_k=b$ for all $k$. When compared against the Poisson--Dirichlet
process, the GDP has quite distinct properties.
%
%
\begin{theorem}\label{thnumclust}
Let $Z_n$ be the number of distinct observations in a sample of size
$n$ from a distribution $G$, where $G \sim\GDP(a,b,G_0)$. The
expected number of clusters $\E(Z_n)$ is given by
%
%
\begin{eqnarray}
\label{eqnumclustexp} \E(Z_n) &=& \sum_{i=1}^{n}
\frac{i a \Gamma(a+b) \Gamma
(b+i-1)}{\Gamma(b)\Gamma(a+b+i) - \Gamma(a+b)\Gamma(b+i)}.
\end{eqnarray}
\end{theorem}

The proof can be seen in Appendix~\ref{approofexpnumc}. Note that for
$a=1$, this expression simplifies to $\E(Z_n) = \sum_{i=1}^{n} \frac
{b}{b + i - 1} \sim o(\log n)$, a well-known result\vspace*{2pt} for the Dirichlet
process [\citet{An74}]. Letting $W_n = Z_n - Z_{n-1}$ denote the change
in the number of clusters in adding the $n$th individual to a sample
with $n-1$ subjects, Stirling's approximation can be used to show that
\[
\E(W_n) = \frac{n a \Gamma(a+b) \Gamma(b+n-1)}{\Gamma(b)\Gamma
(a+b+n) - \Gamma(a+b)\Gamma(b+n)} \approx C(a,b) n^{-a},
\]
where $C(a,b) = \{a\Gamma(a+b)/\Gamma(b) \} \exp\{ -2(a+1)\} $.
Hence, $\E(W_n) \to0$ as \mbox{$n \to\infty$} and new clusters
become increasingly rare as the sample size increases. Note that for $a
\le1$, the number of clusters will grow slowly but without bound as
$n$ increases, with $\E(Z_n) \to\infty$. The rate of growth in
this case is proportional to $n^{1-a}$, which is similar to what is
obtained by using the Poisson Dirichlet prior [\citet{SuJo09}].
However, when $a > 1$ the expected number of clusters instead converges
to a finite constant, which is a remarkable difference compared with
the Dirichlet and Poisson--Dirichlet process. As mentioned above, there
may be a number of practical advantages to bounding the number of
clusters. In addition, a~finite bound on the number clusters seems to
be more realistic in many applications, including the original species
sampling applications that motivated much of the early development in
this area [\citet{MC65,Pi95}].

\begin{figure}

\includegraphics{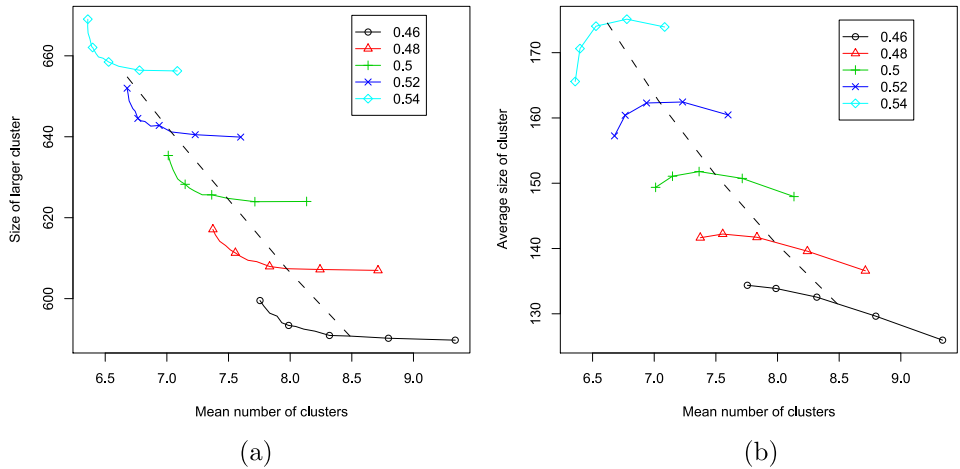}

\caption{Clustering structure induced by a $\GDP(a,b,G_0)$ for a
sample of size $n=1000$. Panel~\textup{(a)} shows the relationship between the
size of the largest cluster and the mean number of clusters for
different GDPs, where each curve shares a common $\E(u_k) = a/(a+b)$.
Panel \textup{(b)} shows the relationship between the average cluster size and
the mean number of clusters. The dashed lines correspond to the
combinations available under a standard Dirichlet process.}\label{ficluststruc}
\end{figure}

In order to gain further insight into the clustering structure induced
by the $\GDP(a,b,G_0)$, we present in Figure~\ref{ficluststruc} the
relationship between the size of the largest cluster and the mean
number of clusters in the partition (left panel), and the mean cluster
size and the number of clusters (right panel) for a sample of size
$n=1000$. The numbers presented in this figure are based on the results
from 20,000 simulations from the stick-breaking prior. Each continuous
line corresponds to a combination of shape parameters such that
$a/(a+b)$ is constant, while the dashed line in the plots corresponds
to the combinations available under a Dirichlet process. The plots
demonstrate that the additional parameter in the GDP allows us to
simultaneously control the number of clusters and the relative size of
the clusters, increasing the flexibility of the model as a clustering procedure.
%
%

The previous discussion focused on the impact of the prior distribution
for the mixture weights on the clustering structure. Another important
issue in the specification of the model is the selection of the
baseline measure $G_0$. Note that in the functional clustering setting
$\bfvartheta_k = (\bftheta^{*}_k, \sigma_k^{*2}, \bfLambda^{*}_k)$
and, therefore, a computationally convenient choice that is in line
with our previous discussion on basis selection and zero-inflated
priors is to write
%
%
\begin{eqnarray}
\label{eqbaselme} G_0\bigl(\bftheta,\sigma^2,\bfLambda
\bigr) &=& \mathsf{N}\bigl( \bftheta| \mathbf{0}, \sigma^2 \bfOmega
\bigr) \times\IGam\bigl(\sigma^2 | \nu_1,\nu_2
\bigr) \times\prod_{s=1}^{p} \ber(
\lambda_s | \gamma).
\end{eqnarray}
A prior of this form allows differential adaptive smoothing for each
cluster in the data; the level of smoothness is controlled by $\gamma$
(the prior probability of inclusion for each of the spline
coefficients) and, therefore, it is convenient to assign to it a
hyperprior such as $\gamma\sim\bet(\eta_1,\eta_2)$.

\section{Functional clustering in nested designs}\label{sefunclustnested}

Consider now the case where multiple curves are collected for each
subject in the study. In this case, the observations consist of ordered
pairs $(y_{ijt}, x_{ijt})$ where
\begin{eqnarray*}
y_{ijt} &=& f_{ij}(x_{ijt}) + \varepsilon_{ijt},
\end{eqnarray*}
where $f_{ij}$ is the $j$th functional replicate for subject $i$, with
$i=1,\ldots, I$, $j=1,\ldots, n_i$ and $t=1,\ldots, T_{ij}$. For
example, in the EPS, $f_{ij}$ is the measurement error-corrected smooth
trajectory in the progesterone metabolite PdG over the $j$th menstrual
cycle from woman $i$, with $t$ indexing the sample number and $x_{ijt}$
denoting the day within the $i,j$ menstrual cycle relative to a marker
of ovulation day.

A natural extension of (\ref{eqsimplmodelh}) to nested designs arises
by modeling the expected evolution of progesterone in time for cycle
$j$ of woman $i$ as $f_{ij} = \bfB(\bfx_{ij}) \bftheta_{ij}$ and
using a hierarchical model for the set of curve-specific parameters $\{
\bftheta_{ij} \}$ in order to borrow information across subjects
and/or replicates. In the following subsections, we introduce two
alternative nonparametric hierarchical priors that avoid parametric
assumptions on the distribution of the basis coefficients, while
inducing hierarchical functional clustering.

\subsection{Mean-curve clustering}\label{semeanclustering}

As a first approach, we consider a Gaussian mixture model, which
characterizes the basis coefficients for functional replicate $j$ from
subject $i$ as conditionally independent draws from a Gaussian
distribution with subject-specific mean and variance, in the spirit of
\citet{BoCaHo08}:
%
%
\begin{eqnarray}\label{eqmeanclustre}
\bfy_{ij} | \bftheta_{ij},
\sigma_i &\sim& \mathsf{N}\bigl( \bfB(\bfx_{ij})
\bftheta_{ij}, \sigma_i^2 \mathbf{I}\bigr),\qquad
\bftheta_{ij} | \bftheta_{i}, \bfLambda_i,
\sigma^2_{i} \sim G_i,
\nonumber\\[-8pt]\\[-8pt]
G_i &=& \mathsf{N}\bigl( \bfLambda_i \bftheta_i,
\sigma^2_{i} \bfSigma\bigr),\nonumber
\end{eqnarray}
where $\bfLambda_i, \bftheta_i, \sigma_i^2$ are as described in
expression (\ref{eqsimplmodelh}) and $\bfSigma$ is a $p \times p$ covariance matrix. In
this model, the average curve for subject $i$ is obtained as $\E\{
f_{ij}(x) | \bfLambda_i, \bftheta_i, \sigma_i^2 \} = {\mathbf
B}(x)\bfLambda_i\bftheta_i$, with $\bfLambda_i$ providing a
mechanism for subject-specific basis selection, so that the curves from
subject $i$ only depend on the basis functions corresponding to nonzero
diagonal elements of $\bfLambda_i$. The variability in the replicate
curves for the same subject is controlled by $\sigma_i^2 \bfSigma$,
with the subject-specific multiplier allowing subjects to vary in the
degree of variability across the replicates. The need to allow such
variability is well justified in the hormone curve application.

In order to borrow information across women, we need a hyperprior for
the woman specific parameters $\{ (\bfLambda_i, \sigma_i^2, \bftheta
_i) \}_{i=1}^I$. Since we are interested in clustering subjects, a
natural approach is to specify this hyperprior nonparametrically
through a generalized Dirichlet process centered around the baseline
measure in (\ref{eqbaselme}), just as we did for the single curve
case. This yields
\begin{eqnarray*}
\bigl(\bftheta_{i}, \sigma_i^2,
\bfLambda_{i}\bigr) | G&\sim& G,\qquad G \sim\GDP( a, b,
G_0)
\end{eqnarray*}
with $G_0$ given in (\ref{eqbaselme}). Since the distribution $G$ is
almost surely discrete, the model identifies clusters of women with
similar average curves. This is clearer if we integrate the
curve-specific coefficients $\{ \bftheta_{ij} \}$ and the unknown
distribution $G$ out of the model to obtain
%
%
\begin{eqnarray}\label{eqjointliksinglecurve}
&& \bfy_{i1},\ldots,\bfy_{in_i} |
\{w_k\}, \bigl\{ \bftheta_k^{*}\bigr\}, \bigl
\{ \sigma_k^{*2}\bigr\}, \bigl\{\bfLambda_k^{*}
\bigr\}
\nonumber\\[-8pt]\\[-8pt]\nonumber
&&\qquad \sim
\sum_{k=1}^{K} w_k \Biggl\{
\prod_{j=1}^{n_i} \mathsf{N} \bigl( \bfB(
\bfx_{ij}) \bfLambda_k^{*} \bftheta_k^{*}, \sigma^{*2}_{k} (\mathbf{I}
+ \bfSigma) \bigr) \Biggr\}.
\end{eqnarray}

By incorporating the distribution of the selection matrices $\bfLambda
_1, \ldots, \bfLambda_I$ in the random distribution $G$, this model
allows for a different smoothing pattern for each cluster of curves.
This is an important difference with a straight generalization of the
model in \citet{RaMa06}, who instead treat the selection matrix as a
hyperparameter in the baseline measure $G_0$ and therefore induce a
common smoothing pattern across all clusters. 

The model is completed by assigning priors for the hyperparameters. For
the random effect variances we take inverse-Wishart priors:
\begin{eqnarray*}
\bolds{\Omega} & \sim& \IWis( \nu_{\Omega}, \bolds{
\Omega}_0 ), \qquad\bolds{\Sigma} \sim\IWis( \nu_{\Sigma},
\bolds{\Sigma}_0 ).
\end{eqnarray*}

In the spirit of the unit information priors [\citet{Pa06}], the
hyperparameters for these priors can be chosen so that $\bolds{\Omega
}_0$ and $\bolds{\Sigma}_0$ are proportional to
\[
\sum_{i=1}^{I} \sum
_{j=1}^{n_i}\mathbf{B}(\bfx_{ij})'
\mathbf{B}(\bfx_{ij}). %
\]
Finally, the concentration parameters $a$ and $b$ are given gamma
priors $a \sim\Gam(\kappa_{a}, \tau_{a})$ and $b \sim\Gam(\kappa
_{b}, \tau_{b})$ and the probability of inclusion $\gamma$ is
assigned a beta prior, $\gamma\sim\bet(\eta_1, \eta_2)$.

\subsection{Distribution-based clustering}\label{sedistclustering}

Because the subject-specific distributions $\{ G_i \}_{i=1}^{I}$ were
assumed to be Gaussian and the nonparametric prior was placed on their
means, the model in the previous section clusters subjects based on
their average profile. However, as we discussed in Section~\ref{seintro}, clustering based on the mean profiles might be misleading in
studies such as the EPS in which there are important differences among
subjects in not only the mean curve but also the distribution about the
mean. In hormone curve applications, it is useful to identify clusters
of trajectories over the menstrual cycle to study variability in the
curves and identify outlying cycles that may have reproductive
dysfunction. It is also useful to cluster women based not simply on the
average curve but on the distribution of curves. With this motivation
in mind, we generalize our hierarchical nonparametric specification to
construct a model with additional parameters that enables clustering
based on more than proximity between the mean curves. This
generalization further enables clustering within subjects as well as subjects.

To motivate our nonparametric construction, consider first the simpler
case in which there are only two types of curves in each cluster of
women (say, normal and abnormal), so that it is natural to model the
subject-specific distribution as a two-component mixture where
%
%
\begin{eqnarray}
\label{seBNPmixfunc0}
&& \bfy_{ij} | \varpi_i, {
\bfLambda}_{1i}, {\bftheta}_{1i}, {\sigma}^2_{1i},
{\bfLambda}_{2i}, {\bftheta}_{2i}, {\sigma}^2_{2i}
\nonumber\\[-8pt]\\[-8pt]\nonumber
&&\qquad \sim
\varpi_i \mathsf{N}\bigl( \bfB(\bfx_{ij}) {
\bfLambda}_{1i} {\bftheta}_{1i}, {\sigma}^2_{1i}
\mathbf{I} \bigr) + (1-\varpi_i) \mathsf{N}\bigl(\bfB(
\bfx_{ij}) {\bfLambda}_{2i} {\bftheta}_{2i}, {
\sigma}^2_{2i} \mathbf{I} \bigr),
\end{eqnarray}
$\varpi_i$ can be interpreted as the proportion of curves from subject
$i$ that are in group 1 (say, normal), and $( {\bftheta}_{1i}, {\sigma
}^2_{1i} )$ are the parameters that describe curves from a normal
cycle, ${\bfLambda}_{1i}$ is a diagonal variable selection matrix for
the normal cycles that zeroes out unnecessary coefficients to avoid
overfitting, $( {\bftheta}_{2i}, {\sigma}^2_{2i} )$ are the
parameters describing the curves from an abnormal cycle, and
${\bfLambda}_{2i}$ is the variable selection matrix for the abnormal
cycles. Note that in this case we have not one but two variance
parameters for each individual, which provide additional flexibility by
allowing each cluster of curves to present a different level of
observational noise. This feature is desirable in the EPS because, for
a given woman, observational noise in abnormal cycles tends to be
larger than in normal cycles.

Under\vspace*{1.5pt} this formulation, the subject-specific distribution is described
by the vector of parameters $(\varpi_i, {\bfLambda}_{1i}, {\bftheta
}_{1i}, {\sigma}^2_{1i}, {\bfLambda}_{2i}, {\bftheta}_{2i}, {\sigma
}^2_{2i})$, and clustering subjects could be accomplished by clustering
these vectors. We can accomplish this by using another mixture model
that mimics (\ref{eqsimplmodel}) and (\ref{eqjointliksinglecurve}),
so that
%
%
\begin{eqnarray}
\label{eqtwocomponent}
&& \bfy_{i1},\ldots,\bfy_{in_i} | \{
\pi_k \}, \{ \varpi_k \}, \bigl\{ \bftheta^{*}_{1k}
\bigr\}, \bigl\{\sigma_{1k}^{*2} \bigr\}, \bigl\{
\bfLambda_{1k}^{*} \bigr\}, \bigl\{ \bftheta^{*}_{2k}
\bigr\}, \bigl\{\sigma_{2k}^{*2} \bigr\}, \bigl\{ \bfLambda
_{2k}^{*} \bigr\} \nonumber
\\
&&\qquad \sim
\sum_{k=1}^{K} \pi_k
\prod_{j=1}^{n_i} \bigl\{
\varpi_k \mathsf{N}\bigl( \bfB(\bfx_{ij}) {
\bfLambda}^{*}_{1k} {\bftheta}^{*}_{1k},
{\sigma}^{*2}_{1k} \mathbf{I} \bigr)
\\
&&\hspace*{80pt}{} + (1-\varpi_k) \mathsf{N}\bigl( \bfB(\bfx_{ij}) {
\bfLambda}^{*}_{2l} {\bftheta}^{*}_{2k}, {\sigma}^{*2}_{1k} \mathbf{I}
\bigr) \bigr\}.\nonumber
\end{eqnarray}

To formulate our Bayesian nonparametric model for clustering, we start
by rewriting (\ref{eqtwocomponent}) as a general mixture model where
%
%
\begin{eqnarray}
\label{eqNDPh} \bfy_{ij} | \bftheta_{ij},
\sigma^2_{ij}, \bfLambda_{ij} & \sim& \mathsf{N}
\bigl( \mathbf{B}(\bfx_{ij})\bfLambda_{ij} \bftheta
_{ij}, \sigma^2_{ij} \mathbf{I} \bigr), \qquad
\bftheta_{ij}, \sigma^2_{ij},
\bfLambda_{ij} | G_i \sim G_i
\end{eqnarray}
and $G_i$ is a discrete distribution which is assigned a nonparametric
prior. Note that this is analogous to the formulation in (\ref
{eqmeanclustre}), but by replacing the Gaussian distribution with a
random distribution with a nonparametric prior we are modeling the
within-subject variability by clustering curves into groups with
homogeneous shape.

Now, we need to define a prior over the collection $\{G_i\}_{i=1}^{I}$
that induces clustering among the distributions. For example, we could
use a discrete distribution \textit{whose atoms are in turn random
distributions}, for example,
\begin{eqnarray*}
G_i & \sim& \sum_{k=1}^{\infty}
\pi_{k} \delta_{ G_k^{*} },
\end{eqnarray*}
where $\pi_{k} = v_k \prod_{s<k} (1-v_s)$, $v_k \sim\bet(a_1, b_1)$
and $G_k^{*} \sim\GDP(a_2, b_2, G_0)$ independently. This implies that
\begin{eqnarray*}
G_k^{*} & =& \sum_{l=1}^{\infty}
\varpi_{lk}\delta_{(\bftheta
_{lk}^{*}, \sigma_{lk}^{2*}, \bfLambda_{lk}^{*})}, \qquad\bigl(\bftheta
_{lk}^{*}, \sigma_{lk}^{2*},
\bfLambda_{lk}^{*}\bigr) \sim G_0,
\end{eqnarray*}
with $\varpi_{lk} = u_{lk} \prod_{s<l} (1-u_{sk})$ and $u_{lk} \sim
\bet(a_2, b_2)$ and $G_0$ as in (\ref{eqbaselme}). Therefore, if we
were to replace the collection $\{G^{*}_k\}_{k=1}^{\infty}$ with
random discrete distributions with only two atoms and we were to
integrate over the random distributions $\{G_i\}_{i=1}^{I}$, this model
would be equivalent to (\ref{eqtwocomponent}) with $K=\infty$.

This model on the collection $\{G_i\}_{i=1}^{I}$ is a generalization of
the nested Dirichlet process introduced in \citet{RoDuGe08a} and, as
with other models based on nested nonparametric processes, interesting
special cases can be obtained by considering the limit of the precision
parameters. For example, letting $b_2 \to0$ while keeping $a_2$ fixed
induces a model where all menstrual cycles within a woman are assumed
to have the same profile, and subjects are clustered according to their
mean cycle. Such a model is equivalent to the one obtained by taking
$\bfSigma\to0$ in (\ref{eqmeanclustre}). On the other hand, by
letting $b_1 \to\infty$ while keeping $a_1$ constant, we obtain a
model where all subjects are treated as different and menstrual cycles
are clustered within each women. In this case, information is borrowed
across the menstrual cycles of each women, but not across women.

Intuitively, we can think of the model we just described as first
clustering curves within a subject via a model-based version of $k$-means
applied to the subject-specific basis coefficients. Then, two subjects
having similar distributions of curve clusters will be clustered
together. Because we use a joint hierarchical model, these stages are
done simultaneously.


Again, the model is completed by specifying prior distributions on the
remaining parameters. As before,
we let $\bfOmega\sim\IWis( \nu_{\Omega}, \bfOmega_0)$, $\nu_{2}
\sim\Gam( \rho, \psi)$ and $\gamma\sim\bet(\eta_1,\eta_2)$,
providing a conditionally conjugate specification amenable for simple
computational implementation. Finally, for the precision priors of the
GDPs we set
\begin{eqnarray*}
a_1 &\sim& \Gam(\kappa_{a_1},\tau_{a_1}),\qquad
b_1 \sim\Gam(\kappa_{b_1},\tau_{b_1}),
\\
a_2 &\sim& \Gam(\kappa_{a_2},\tau_{a_2}),\qquad
b_2 \sim\Gam(\kappa_{b_2},\tau_{b_2}).
\end{eqnarray*}

\section{Computation}\label{secompu}

As is commonplace in Bayesian inference, we resort to Markov chain
Monte Carlo (MCMC) algorithms [\citet{RoCa99}] for computation in our
functional clustering models. Given an initial guess for all unknown
parameters in the model, the algorithms proceed by sequentially
sampling blocks of parameters from their full conditional
distributions. In particular, we design our algorithms using truncated
versions of the GDP and the nested GDP, where a large but finite number
of atoms are used to approximate the nonparametric mixture
distributions. Well-known results on the convergence of truncations as
the number of atoms grows that were originally presented in \citet
{IsJa01} and \citet{RoDuGe08a} can be directly extended to this problem
(see Appendix~\ref{aptruncGDP}). Furthermore, because most components
of the model are conditionally conjugate, most of the full conditional
distributions can be directly sampled using Gibbs steps. Full details
of the algorithm can be seen in the online supplementary materials [\citet{supp20014}].

Convergence of the MCMC algorithms was assessed using the multi-chain
method described in \citet{GeRu92}, which was applied to monitor the
(unnormalized) posterior distribution, as well as the number of
occupied clusters on each level of the model and the Frobenius norm of
the matrices $\bfOmega$ and $\bfSigma$.

\section{The Early Pregnancy Study}\label{seillus}

Progesterone plays a crucial role in controlling different aspects of
reproductive function in women, from fertilization to early development
and implantation. Therefore, understanding the variability of hormonal
profiles across the menstrual cycle and across subjects is important in
understanding mechanisms of infertility and early pregnancy loss, as
well as for \mbox{developing} approaches for identifying abnormal menstrual
cycles and women for diagnostic purposes. Our data, extracted from the
Early Pregnancy Study [\citet{WiWeOC88}], consists of log daily
creatinine-corrected concentrations of pregnanediol-3-glucuronide (PdG)
for 60 women along multiple menstrual cycles, measured in micrograms
per milligram of creatinine (\textmu g$/$ml Cr). We focus on \mbox{13-}day
intervals extending from 10 days before ovulation to 2 days after
ovulation. According to the results in \citet{DuWeWiBa99}, this
interval should include the fertile window of the menstrual cycle
during which noncontracepting intercourse has a nonneglible
probability of resulting in a conception. Also, for this illustration
we considered only nonconceptive cycles and women with at least four
cycles in each record. Therefore, the number of curves per woman varies
between 4 and 9. 

We analyzed the EPS data using both the mean-based clustering model
described in Section~\ref{semeanclustering} and the distribution-based
clustering model of Section~\ref{sedistclustering}. These models where
fitted using the algorithms from Section~\ref{secompu}. In the
mean-based clustering algorithm, the GDP was truncated so that $K=40$,
while in the distribution-based algorithm the nested GDP was truncated
so that $K=40$ and $L=30$. Although these numbers might seem large
given the sample sizes involved, a large number of empty components are
helpful in improving the mixing of the algorithms. In both cases, we
used piecewise linear splines ($q=1$) and $p=13$ knots, corresponding
to each of the days considered in the study. All inferences presented
in this section are based on 100,000 samples obtained after a burn-in
period of 10,000 iterations.

Prior distributions in the mean-based clustering algorithm were set as
follows. For the concentration parameters, we used proper priors $a
\sim\Gam(3,3)$ and $b \sim\Gam(3,3)$, and for the observational
variance, we set $\sigma^2 \sim\IGam(2,0.04)$, so that $\E(\sigma
^2) = 0.04$. Note that setting $\E(a) = 1$ a priori is natural because
(as we discussed in Section~\ref{seBNPmixfunc}) the asymptotic
behavior of the expected number of clusters is different if $a > 1$ or
$a \le1$. On the other hand, the prior mean for $\sigma^2$ is based
on information from previous studies. To allow uncertainty in the
probability of basis selection within the base measure, we let $\gamma
\sim\bet(2,4)$, implying that we expect about one-third of the spline
basis functions to be used in any given cluster. Priors for the
distribution-based clustering algorithm were chosen in a similar way,
with $a_1 \sim\Gam(3,3)$, $b_1 \sim\Gam(3,3)$, $a_2 \sim\Gam
(3,3)$ and $b_2 \sim\Gam(3,3)$, while for the baseline measure we
picked a prior with the inclusion probabilities $\gamma\sim\bet
(2,4)$ and the prior on the group specific variances as given by $\IGam
(2,0.04)$.

%
%
\begin{figure}

\includegraphics{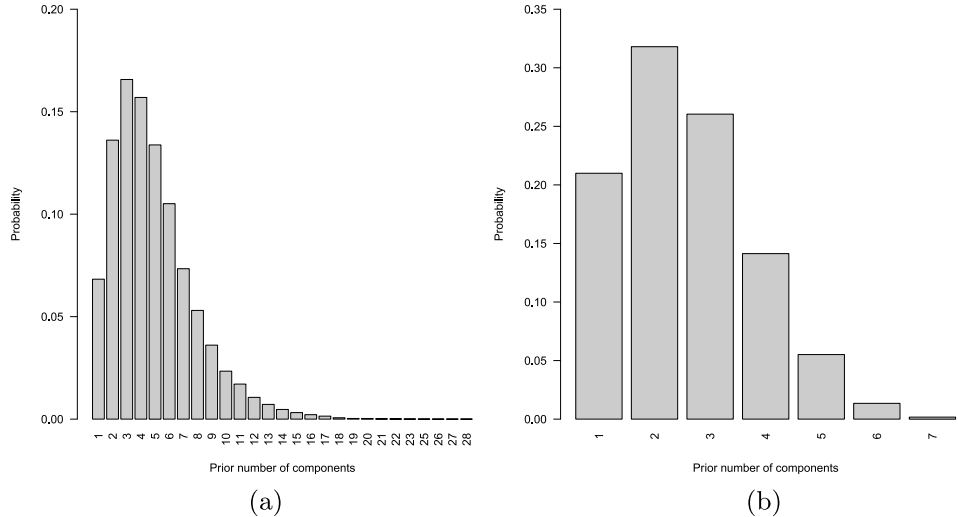}

\caption{Prior number of clusters implied by our specification of $a$,
$b$, $a_1$, $b_1$, $a_2$ and $b_2$.
\textup{(a)}~$n=60$ (subject-level clustering),
\textup{(b)}~$n=7$  (curve-level clustering).}\label{fipriorcomp}
\end{figure}

To better understand the effect of our prior on $a$, $b$, $a_1$, $b_1$,
$a_2$ and $b_2$, we show in Figure~\ref{fipriorcomp} the prior
expected number of clusters implied by our choices. In particular,
Figure~\ref{fipriorcomp}(a) shows that, a priori, we expect between 1
and 8 clusters of subjects with high probability, with the most likely
prior value being 3 clusters (note that this applies to both the
mean-based and the distribution-based clustering methods). On the other
hand, Figure~\ref{fipriorcomp}(b) shows that, for a subject for which
7 cycles have been observed, we expect between 1 and 3 clusters with
high probability. On the other hand, to assess the effect of the prior
choices on the results, we conducted a small sensitivity analysis. In
particular, the priors for $a$, $b$, $a_1$, $a_2$, $b_1$ and $b_2$ were
replaced with exponential distributions with mean 2. The induced priors
on the number of clusters have similar modes to our original
specifications but have higher variability, placing substantially more
mass on larger number of clusters. Although the posterior distribution
for these parameters was somewhat affected, we did not see any
substantial change in our posterior inferences on the clustering
structure or the curve shapes (which are the main focus of our
analysis). Similarly, we explored the effect of a $\bet(1,1)$ prior on
$\gamma$ and an $\IGam(2, 0.1)$ prior on $\sigma^2$ without any
significant change in the estimates of the hormonal profiles.

\begin{figure}

\includegraphics{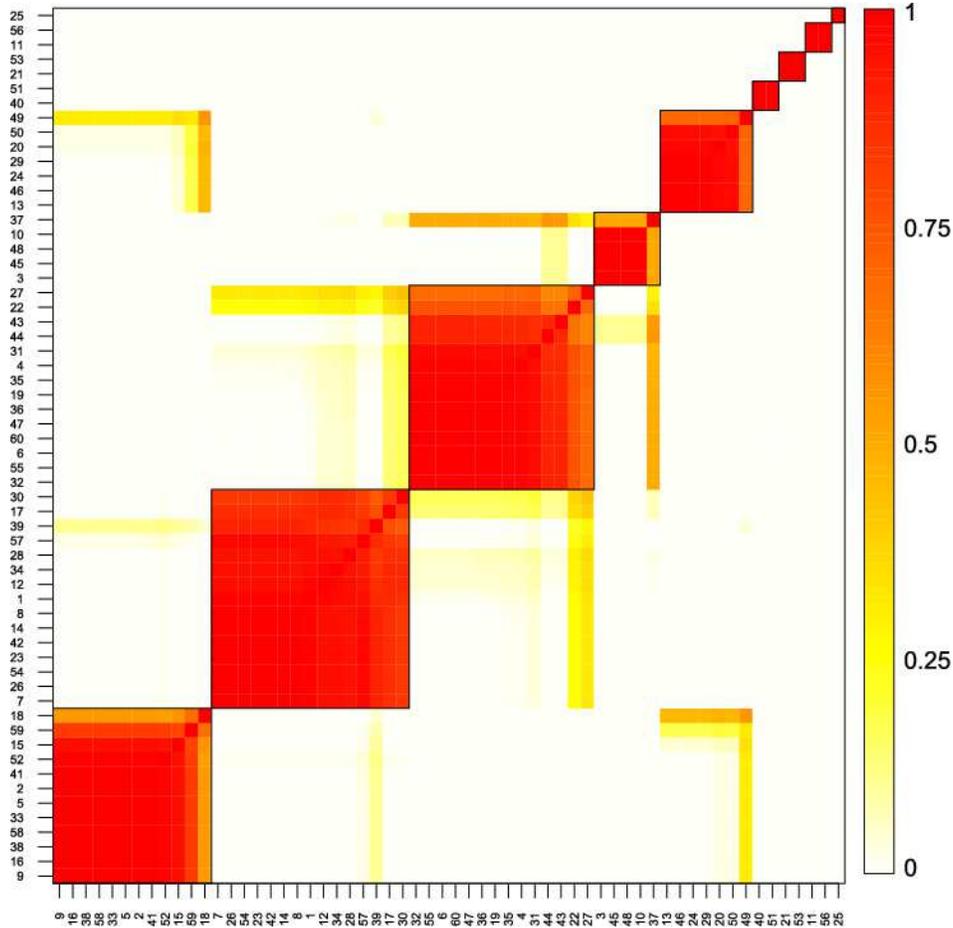}

\caption{Average incidence matrix, illustrating probabilities of joint
pairwise classification for the 60 women in the EPS under the
mean-curve clustering procedure described in Section~\protect\ref
{semeanclustering}. White corresponds to zero probability, while red
corresponds to 1. The squares correspond to a point estimate of the
cluster structure in the data.}\label{fipairhormeanclust}
\end{figure}

\begin{figure}

\includegraphics{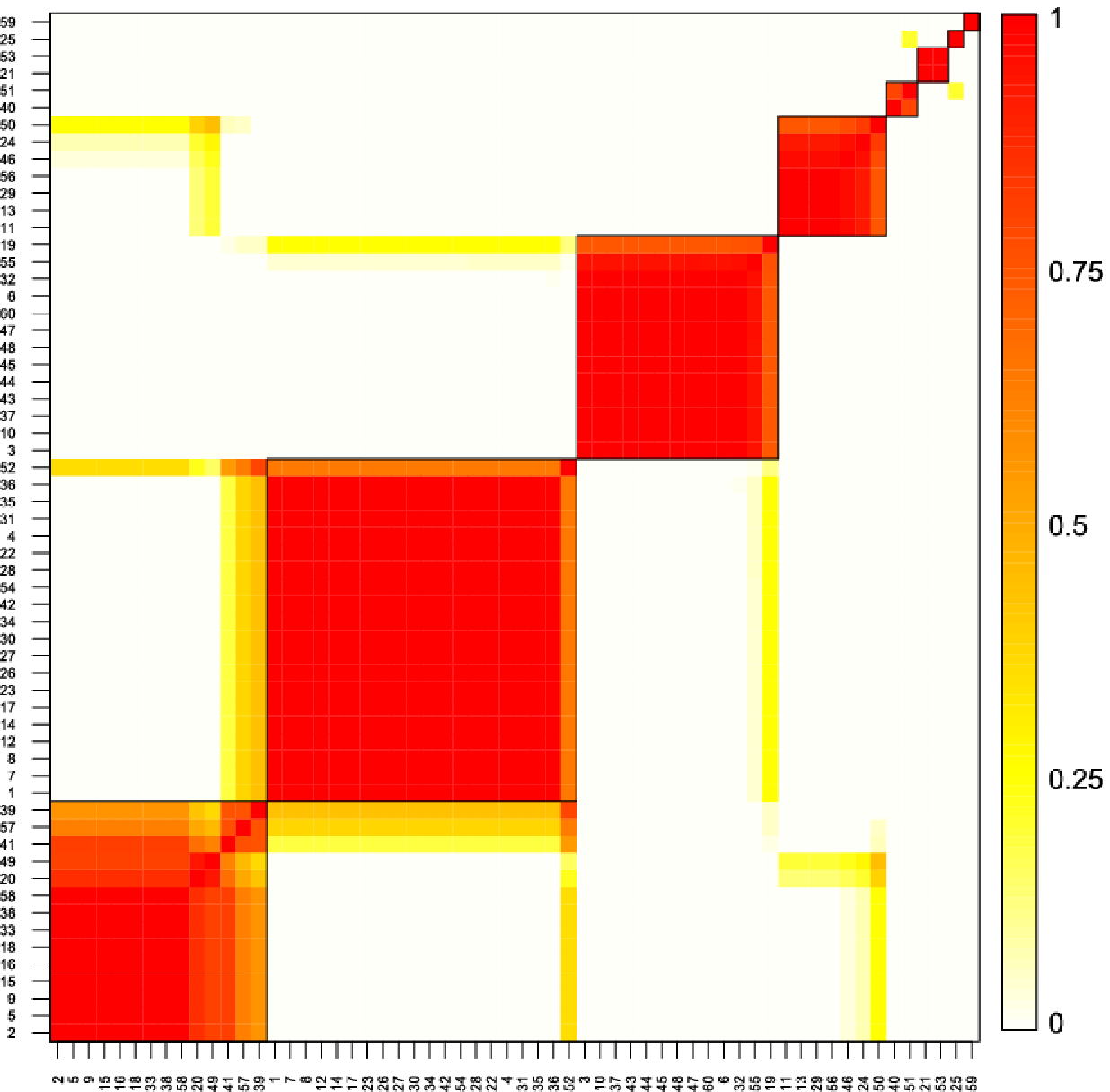}

\caption{Average incidence matrix, illustrating probabilities of joint
pairwise classification for the 60 women in the EPS under the
distribution-based clustering procedure described in Section~\protect
\ref{sedistclustering}. White corresponds to zero probability, while
red corresponds to 1. The squares correspond to a point estimate of the
cluster structure in the data.}\label{fipairhordistclust}
\end{figure}

We start our analysis by comparing the clustering structure generated
by the mean-based and distribution-based models considered in
Section~\ref{sefunclustnested}. For this purpose, we show in
Figures~\ref{fipairhormeanclust} and \ref{fipairhordistclust}
heatmaps of the average pairwise clustering probability matrix under
these two models. Entry $(i,j)$ of the matrix contains the posterior
probability that observations $i$ and $j$ are assigned to the same
cluster. The black squares in the plots correspond to point estimates
of the clustering structure obtained through the method described in
\citet{LaGr06}. In our case, the point estimate is obtained by
minimizing a loss function that assigns equal weights to all pairwise
misclassification errors. Therefore, the resulting plots provide
information about the optimal clustering structure for the data as well
as the uncertainty associated with it.

\begin{figure}

\includegraphics{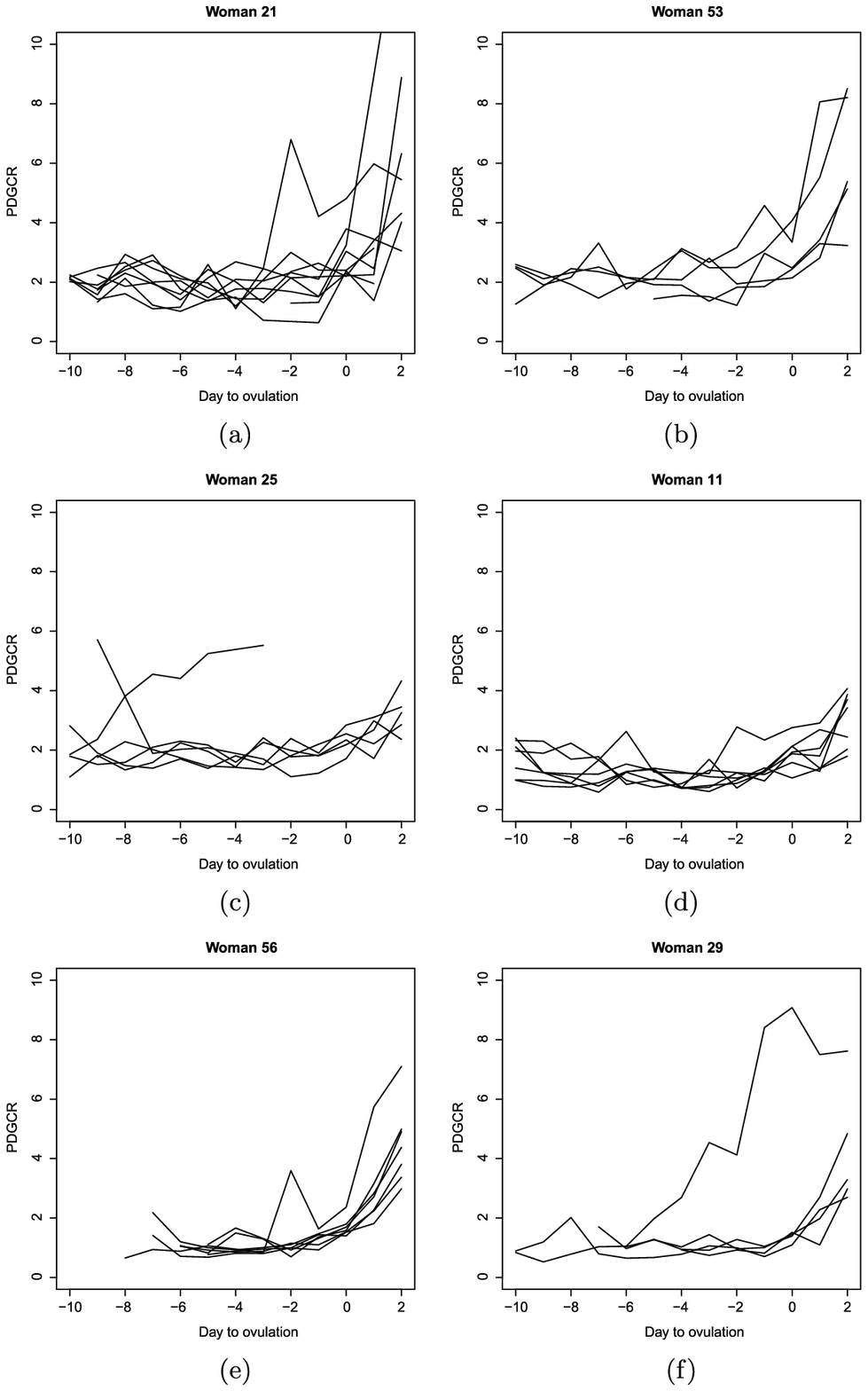}

\caption{Raw data associated with the hormonal profiles for three
outlier subjects in the Early Pregnancy Study.}\label{fihorraw2}
\end{figure}

Figures~\ref{fipairhormeanclust} and \ref{fipairhordistclust} show that the structure of the clusters generated by
both models are similar. For example, cluster 1 from mean-based
clustering (counting from the bottom left corner of Figure~\ref{fipairhormeanclust}) contains
similar subjects as cluster 1 from distribution-based clustering
(counting from the bottom left corner of Figure~\ref{fipairhordistclust}). The same is true
for cluster 2 from both algorithms, and for cluster 5 from the
mean-based clustering model and cluster 4 from the distribution-based
model. There are also similarities in the structure of the outlier
clusters (corresponding to the small clusters at the top and right of
both heat maps). Both models assign subject 25 to a singleton cluster,
while grouping subjects 21 and 53 into a pair, with subjects 40 and 51
in a separate pair. There are also some important differences between
the two methods. Mean-based clustering groups subjects 36 and 43
together and assigns subject 3 to a different cluster (Figure~\ref{fihorpro} shows
differences in the mean curve of subject 3), while distribution-based
cluster groups subjects 3 and 43 together and assigns 36 to a different
group. The behavior illustrates the robustness of distribution-based
clustering, because except for a single outlier cycle for subject 43,
the curves for this subject are much more similar to those of subject 3
than those of 36. Mean-based clustering also treats subjects 11 and 56
as singletons, while distribution-based clustering assigns them to the
large cluster 4 (counting from the bottom left). Visually the raw
profiles for subjects 11 and 56 are not very different from other
profiles in cluster 4 [see panels (d), (e) and (f) of Figure~\ref{fihorraw2}], so
this behavior seems reasonable.
%
%
%
%
%
%

\begin{figure}

\includegraphics{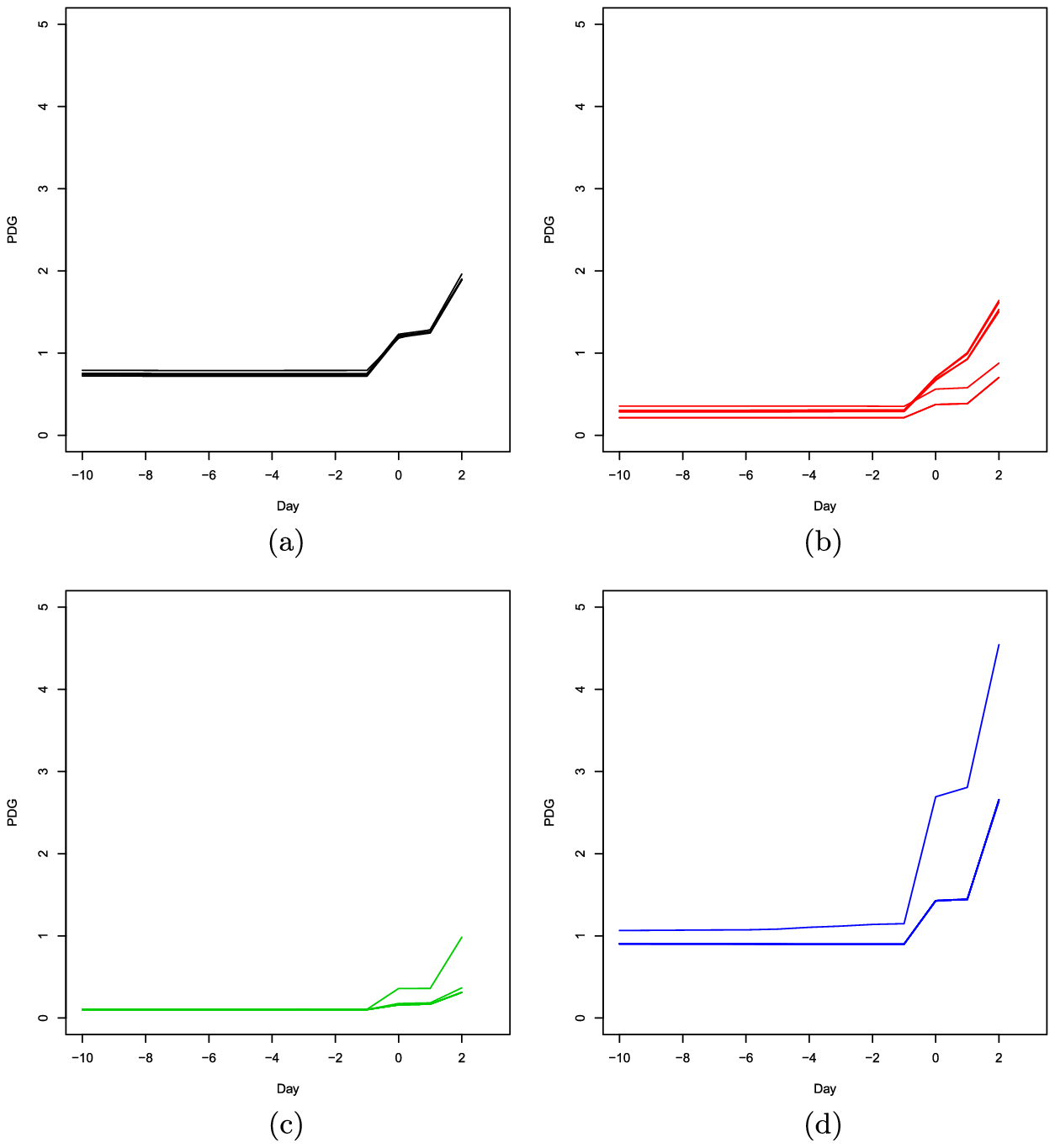}

\caption{Reconstructed profiles for some representative subjects in
the study. Panel \textup{(a)} corresponds to patient 9 (who was chosen from
cluster 1, counting from the bottom left), panel \textup{(b)} to patient 36 (who
was chosen from cluster 2), panel \textup{(c)} to patient 45 (who was chosen
from cluster 3), and panel~\textup{(d)} corresponds to patient 13 (who was
chosen from cluster 4).}\label{firechor}
\end{figure}

Figure~\ref{firechor} shows reconstructed profiles under the
distribution-based clustering model for some representative women in
each of the main four groups. Most profiles are flat before ovulation,
when hormone levels start to increase. Also, in most clusters the
profiles tend to be relatively consistent for any single woman.
However, we can see some outliers, typically corresponding to elevated
post-ovulation levels and/or early increases in the hormone levels.
Cluster 3 corresponds to women with very low hormonal levels, even
after ovulation. This group has few outliers, and those present are
characterized by a slightly larger increase in progesterone after
ovulation, which is still under 1~\textmu g$/$ml Cr. Group 2 shows much
more diversity in the hormonal profiles, as well as a slightly higher
baseline level in progesterone level and an earlier rise in
progesterone than group 3. Group 1 tends to show few outliers, and
otherwise differs from the previous ones in a higher baseline level and
an early and very fast increase in progesterone. Finally, group 4
presents ``normal'' cycles with the highest baseline level of
progesterone (1~\textmu g$/$ml Cr) and the fastest increase in progesterone
after ovulation, along with ``abnormal'' cycles with even higher
baseline levels and very extreme levels of progesterone after ovulation
(close to 5~\textmu g$/$ml Cr).
%
%

%
%
\begin{figure}

\includegraphics{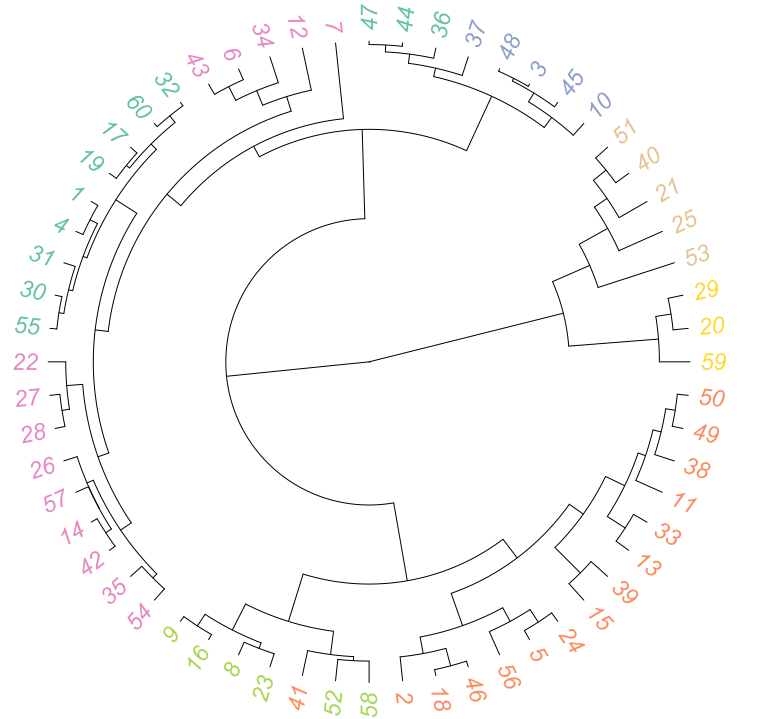}

\caption{Dendogram for an alternative functional clustering algorithm
that uses kernel smoothing to project all curves to a common set of
grid knots and then applies complete-linkage hierarchical clustering
and BIC to create a partition of the data.}\label{fidendogram}
\end{figure}

For comparison we also applied a simple functional clustering approach
based on kernel smoothing and hierarchical clustering. We first fitted
a Gaussian kernel smoother [\citet{LiRa04,RaLi04}] to each of the
curves, generated fitted values over a common grid (in this example we
used 6 equally-spaced knots), then computed the average predicted
values for each subject at each point of the grid, and finally applied
complete-linkage hierarchical clustering (as implemented in the \texttt{R} package \texttt{mclust}), with the number of clusters selected
using~BIC. This approach identified 7 clusters; Figure~\ref
{fidendogram} shows the associated dendogram, where colors are used to
represent the clusters. Some of these clusters are similar to the ones
identified by the mean-based clustering model. For example, the small
cluster of 5 subjects (3, 10, 37, 45 and 48), corresponding to the
fourth cluster from the bottom left in Figure~\ref{fipriorcomp}, is perfectly
identified by hierarchical clustering. Similarly, subjects 21, 25, 40,
51 and 53, which are all identified as outliers by the mean-based
clustering model, are allocated to a single cluster by the hierarchical
clustering approach. However, the majority of the clustering pattern
generated by this method is quite different from the one obtained with
mean-based or the distribution-based clustering. Furthermore, the
number and structure of clusters depends heavily on the number and
location of grid points used to interpolate the functions. For example,
when the interpolation grid contains 5 knots we obtain 8 clusters,
while 18 clusters are obtained when 11 knots are used for
interpolation. Part of this difference could be due to the fact that,
because most of the missing values tend to concentrate at the beginning
of the cycle, many of the fitted functional values correspond to
extrapolations rather than interpolations when a large number of knots
are used.

We also note that posterior estimates of the precision parameters on
the GDP suggest that a logarithmic rate of growth for the expected
number of clusters might be reasonable for this data. For the
mean-based clustering, the posterior mean for $a$ was $1.06$ and the
95\% posterior symmetric credible interval was $(0.65, 1.49)$, while
the posterior mean for $b$ was $0.77$ with 95\% credible interval
$(0.17, 1.72)$. For the distribution-based clustering, the
corresponding estimates are $1.03$ $(0.62,1.56)$ and $0.72$
$(0.20,1.73)$ for $a_1$ and $b_1$, and $1.12$ $(0.71,1.43)$ and $0.27$
$(0.15,1.51)$ for $a_2$ and $b_2$.

\begin{figure}

\includegraphics{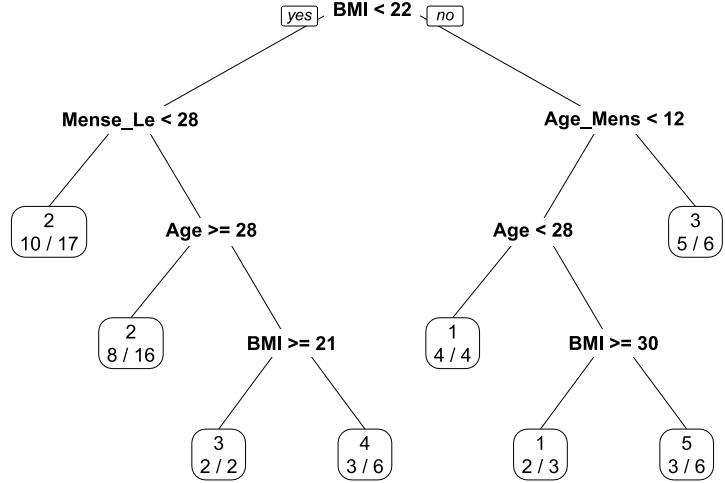}

\caption{Classification tree explaining cluster membership on the
basis of six covariates: age at the beginning of the study (Age), age
of first menses (Age\_Mense), average length of menses (Mense\_Length),
race (Race), body mass index (BMI), whether the subject had been taking
contraceptives before the start of the study (EV\_Pill) and whether the
subject self reports as a consumer of marijuana during the period of
the study (Marijuana\_Use).}\label{ficlasstree}
\end{figure}

Finally, we investigated the relationship between the clusters we
identified using our distribution-based clustering algorithm and a
series of covariates available for each subject, including age at the
beginning of the study (Age), age of first menses (Age\_Mense), average
length of menses (Mense\_Length), race (Race), body mass index (BMI),
whether the subject had been taking contraceptives before the start of
the study (EV\_Pill) and whether the subject self reports as a consumer
of marijuana during the period of the study (Marijuana\_Use). To
investigate this relationship, we used a classification tree, as
implemented in the \texttt{R} package \texttt{rpart}. To simplify
interpretation, we merged all the small outlier clusters into a single
group (which we label cluster 5). The resulting classification tree
(obtained using a Cp value of 0.03 for pruning) can be seen in
Figure~\ref{ficlasstree}. Note that race and marijuana consumption do
not appear to play any role in predicting the shape of the hormonal
profiles. On the other hand, BMI, along with age, menses length and
menses age, seems to have some predictive capability. For example, a
BMI higher than 22 and an age of menses under 12 years old seem to
predict progesterone profiles like the ones observed in cluster 3.

\section{Discussion}\label{seconclusion}

We have presented two approaches to functional clustering in nested
designs. These approaches look into different features of the nested
samples and are therefore applicable in different circumstances. Our
mean-based clustering approach is easier to interpret and provides an
excellent alternative when within-subject samples are homogenous.
However, when within-subject curves are heterogeneous, mean-based
clustering can lead to biased results. Therefore, in studies such as
the EPS, distribution-based models such as the one described here
provide a viable alternative that acknowledges the heterogeneity in the
function replicates from a subject.

%
%

One interesting insight that can be gathered from the results of the
EPS data is that, for small numbers of functional replicates per
subject and rare outliers, the effect of the distribution-based
clustering is to perform clustering based on the modal rather than the
mean profile. That is, the distribution-based clustering model is able
to automatically discount the abnormal curves, leading to more
appropriate clustering patterns if the effect of outliers needs to be
removed. Naturally, this perceived advantage of the distribution-based
clustering method implicitly assumes that abnormal curves should be
discounted. Although that assumption is justified in our application,
users should be aware of it when applying our method to other data sets.

\begin{appendix}
\section{Proof of Theorem 1}\label{approofexpnumc}

Let $\bftheta_1^{*}, \bftheta_2^{*}, \ldots$ be a sequence of
independent and identically distributed samples from a random
distribution $G$, which follows a $\GDP(a,b,G_0)$ distribution. Also,
let $W_i$ be 1 if $\bftheta^{*}_i$ is different from every $\bftheta
_1^{*},\ldots,\bftheta^{*}_{i-1}$, and zero otherwise. Clearly, $Z_n
= \sum_{i=1}^n W_i$ is the number of distinct values among the first
$n$ samples form a $\GDP(a,b,G_0)$. \citet{Hj00} shows that
\begin{eqnarray*}
\E(W_i) & =& i\frac{\E\{u (1-u)^{i-1}\}}{1 - \E\{ (1-u)^{i} \}}
\\
& =&i\frac{((\Gamma(a+b))/(\Gamma(a) \Gamma(b)))
((\Gamma(a+1)\Gamma(b+i-1))/(\Gamma(a+b+i)))}
{
1 - ((\Gamma(a+b))/(\Gamma(a) \Gamma(b)))
((\Gamma(a)\Gamma(b+i))(\Gamma(a+b+i)))}
\\
& =& \frac{i a \Gamma(a+b)\Gamma(b+i-1)}{\Gamma(b)\Gamma(a+b+i) -
\Gamma(a+b)\Gamma(b+i)},
\end{eqnarray*}
which completes the proof.

\section{Truncations of generalized Dirichlet~processes}\label{aptruncGDP}
%
%
\begin{theorem}\label{thbounds}
Assume that samples of $n$ observations have been collected for each of
$J$ distributions and are contained in vector ${\mathbf y} = ({\mathbf
y}_1',\ldots,{\mathbf y}_J')$. Also, let
\begin{eqnarray*}
P^{\infty\infty}(\bftheta) &=& \int\!\!\int P(\bftheta| G_j)
P^{\infty
}(dG_j | Q) P^{\infty}(dQ),
\\
P^{LK}(\bftheta) &=& \int\!\!\int P(\bftheta| G_j)
P^{L}(dG_j | Q) P^{K}(dQ)
\end{eqnarray*}
be, respectively, the prior distribution of the model parameters under
the nested GDP model and its corresponding truncation\vspace*{1pt} after integrating
out the random distributions, and $P^{\infty\infty}({\mathbf y})$ and
$P^{LK}({\mathbf y})$ be the prior predictive distribution of the
observations derived from these priors. Then
\[
\int\bigl\llvert P^{LK}({\mathbf y}) - P^{\infty\infty}({\mathbf y}) \bigr
\rrvert\, d{\mathbf y} \le\int\bigl\llvert P^{LK}(\bftheta) -
P^{\infty\infty
}(\bftheta) \bigr\rrvert\le\varepsilon^{LK}(\alpha,\beta),
\]
where
\begin{eqnarray*}
\varepsilon^{LK}(\alpha,\beta) &=& 
\cases{ \displaystyle4
\biggl( 1 - \biggl[ 1 - \biggl( \frac{b_1}{a_1+b_1} \biggr)^{K-1}
\biggr]^{J} \biggr),\hspace*{25pt}\mbox{if }L =\infty, K < \infty,
\vspace*{8pt}\cr
\displaystyle4 \biggl( 1 - \biggl[ 1 - \biggl( \frac{b_2}{a_2+b_2}
\biggr)^{L-1} \biggr]^{nJ} \biggr),\qquad\mbox{if }L < \infty, K=\infty,
\vspace*{8pt}\cr
\displaystyle4 \biggl( 1 - \biggl[ 1 - \biggl( \frac{b_1}{a_1+b_1}
\biggr)^{K-1} \biggr]^{J} \biggl[ 1 - \biggl(
\frac{b_2}{a_2+b_2} \biggr)^{L-1} \biggr]^{nJ} \biggr),
\vspace*{5pt}\cr
\hspace*{170pt}\mbox{if }L < \infty, K < \infty.}
\end{eqnarray*}
\end{theorem}

The proof is a direct extension of results in \citeauthor{IsJa01} (\citeyear{IsJa01,IsJa02})
and \citet{RoDuGe08a} and it is omitted for
reasons of space.
This result is particularly important since it justifies the use of
computational algorithms based on finite mixtures and allows us to
choose adequate truncation levels.
\end{appendix}

\section*{Acknowledgments}
We would like to thank Nils Hjort for his helpful comments and Allen
Wilcox for providing access to the EPS data.

\begin{supplement}[id=suppA]
\stitle{Supplement to ``Functional clustering in nested designs: Modeling variability in reproductive epidemiology studies''\\}
\slink[doi]{10.1214/14-AOAS751SUPP} 
\sdatatype{.pdf}
\sfilename{aoas751\_supp.pdf}
\sdescription{The supplementary materials contain the details
of the Markov chain Monte Carlo algorithm
used to fit the models introduced in the paper.}
\end{supplement}

%

\printaddresses
\end{document}